\documentclass[a4paper] {article}
\usepackage{graphics,color,epsfig}
\def\linadj#1{\normalbaselines
        \multiply\lineskip#1 \divide\lineskip100
        \multiply\baselineskip#1 \divide\baselineskip100
        \multiply\lineskiplimit#1 \divide\lineskiplimit100 }

\begin{document}

\title{ \bf Speed of sound in a Quark-Gluon-Plasma with one
loop correction in mean-field potential}

\author{S. Somorendro Singh\footnote{email:sssingh@physics.du.ac.in} and R. Ramanathan }
        
\maketitle

\begin{center}
Department of Physics and Astrophysics, University of Delhi, Delhi - 110007, India \\
\end{center}
\linadj{200} 

\begin{abstract}
 We study thermodynamic properties and speed of sound in a 
free energy evolution of
quark-gluon plasma (QGP)
with one loop correction factor in the mean-field potential.
The values of the thermodynamic properties like pressure,
entropy and specific heat are calculated for
a range of temperatures. The results agree with the recent lattice results. 
The speed of sound
is found to be $C_{s}^2=0.3$ independent of parameters used in the
loop correction which matches almost with lattice calculations.
\\
\\
PACS number(s): 25.75 Ld, 12.38 Mh \\
Keywords: Quark-Gluon Plasma, Sound Speed; 
\end{abstract}
\vfill
\eject
\maketitle

\section{Introduction}
\large
\par   The matter formation during the early stage of universe expansion
is considered to be a complicated phenomena. To solve this problem
there are a number of experimental laboratories set up in the last
two decades. It is believed that there is a phase transition during
the early universe expansion to a phase of confining matter of
hadrons from deconfined matter of free quarks and gluons~\cite{karsch,csernai}.
The process of transition is investigated in the experiments like
relativistic heavy-ion collider (RHIC) at BNL and 
large hadron collider (LHC) at
CERN. The study makes
it attractive for the experimentalists and theorists to collaborate in
 this 
 exciting field in the present day of
heavy ion collider physics~\cite{satz,mustafa}.  In this short paper, 
we focus 
on the QGP
evolution through the free energy expansion of the system with one loop
correction. The free energy is constructed through a density of state which
is obtained with a one loop corrected 
mean-field potential. 
So the free energy contribution from every
constituent particles of the system is considered in this calculation.
Moreover, the correction factor in the potential is obtained through
coupling value,~\cite{peshier,raman,singh} and due to
this correction value, there are changes in the 
free energy expansion of QGP fireball, and it also impacts in 
the stability of
droplet with the variation of dynamical quark and gluon 
flow parameters~$(\gamma_{q},\gamma_{g})$.
\par In this short article we calculate thermodynamic parameters
like pressure, entropy and specific heat through the free energy. 
Then we calculate
the speed of sound in a QGP with one loop correction in mean field
potential which is extensively dealt with in our earlier paper~\cite{somoren}.
The construction of density of
states and the free energy with the loop correction has been 
explained in detail with these papers.
In conclusion, 
we give the
details of thermodynamic properties of QGP fireball like pressure, entropy,
specific heat and speed of sound with
the flow parameter $\gamma_{q}=8, \gamma_{g}=8 \gamma_{q}$.

\section{Density of states and free energy}

\par The density of states in phase space with loop correction
in the interacting potential is obtained through a generalized Thomas- Fermi
model as~\cite{ramanathan,fermi}:
\begin{equation}
\int \rho_{q,g}dq=\nu/\pi^{2}[-V_{conf}(q)]^{2} \frac{dV_{conf}}{dq}~,
\end{equation}
where,
\begin{equation}\label{3.18}
V_{\mbox{conf}}(q) = \frac{2}{q}\sqrt{(1/\gamma_{g})^{2} + (1 / \gamma_{q})^{2}} ~ g^{2}(q) T^{2} [1+g^{2}(q) a_{1} ]- \frac{m_{0}^{2}}{2 q}.
\end{equation}
 
$ \gamma_{q}=1/8 $ and $\gamma_{g}=~ (8 - 10)~ \gamma_{q}$ are quark and gluon 
flow parameters chosen in fitting the evolution of 
QGP droplets~\cite{somoren,brambilla,melnikov}.
 Then we finally get the density of state as:
\begin{equation}\label{3.13}
\rho_{q, g} (q) =\frac{\nu}{\pi^2}[\frac{\gamma_{q,g}^{3}T^2}{2}]^{3} g^{6}(q)A, 
\end{equation}
where
\begin{equation}
A=\lbrace 1+\frac{g^2(q)a_{1}}{4\pi^2}\rbrace^{2}[ \frac{(1+g^2(q)a_{1}/(4\pi^2))}{q^{4}}+\frac{ (2+g^2(q)a_{1}/\pi^2)}{q^{2}(q^2+\Lambda^2)\ln(1+\frac{q^2}{\Lambda^2})}] 
\end{equation}
and~ $\nu$ is the volume occupied by the QGP. $q$ is
the relativistic four-momentum in natural units.
$~g^{2}(q)=4 \pi \alpha_{s}(q)$
where,
$~\alpha_{s}(q) $ is the coupling
value of quark and gluon with degree of
freedom~ $n_{f},$ defined as
\begin{equation}
 \alpha_{s}(q)=\frac{4 \pi}{(33-2n_{f})\ln(1+q^{2}/\Lambda^{2})},
\end{equation}
in which~$\Lambda$ is QCD parameter measured in the scale
of lattice QCD and it is taken as~$ 0.15~$GeV.
The coefficient $a_{1}$  in the confining potential of
the above expression
is one loop
correction in their interactions and it is given as~\cite{fischler}:
\begin{equation}
a_{1}= 2.5833-0.2778~ n_{l},~ 
\end{equation}
where $n_{l}$ is the number of light quark elements~\cite{smirnov}.
\par Now we calculate free energy for the system using
the expression as in~\cite{neergaad,mardor}:
\begin{equation}\label{3.20}
F_i = \mp T g_i \int dq \rho_{q,g} (q) \ln (1 \pm e^{-(\sqrt{m_{i}^2 + q^2}) /T})~,
\end{equation}

with minimum energy cut off as:
\begin{equation}\label{3.19}
V(q_{min})=(\gamma_{g,q}N^{\frac{1}{3}} T^{2} \Lambda^4 / 2)^{1/4},
\end{equation}
where $N=(4/3 )[12 \pi / (33-2 n_{f})]$.
The minimum cut off in
the model makes the integral to have a finite value by avoiding
the infra-red divergence while taking
the magnitude of $\Lambda$ and
$T$ as of the same order as in lattice QCD.
 $g_{i}$ is the corresponding degeneracy factors for the subsequent 
free energies.
 The inter-facial energy
obtained through a scalar
Weyl-surface in Ramanathan et al. \cite{ramanathan,weyl} with a suitable
modification to take care of
the hydrodynamic effects is given as:
\begin{equation}\label{3.22}
  F_{interface}= \frac{1}{4}\sqrt(1/\gamma_{q}^2+1/\gamma_{g}^2) R^{2}T^{3}. 
\end{equation}
The energy is used in replacing the role of bag energy of MIT model 
and it minimizes
the drawback produced in numerical techniques by MIT model.  
The hadronic free energy is~\cite{balian}

\begin{equation}\label{3.25}
F_{h} = (g_{i}T/2\pi^2 )\nu \int_0^{\infty} q^2 dq \ln (1 - e^{-\sqrt{m_{h}^2 + q^2} / T}).
\end{equation}
For quark free energy we consider
quark masses $m_u = m_d = 0 ~ MeV$
and $m_s = 0.15 ~ GeV$~ [15]. The hadronic particles are taken having the
masses below $1.5~$GeV and their contribution to the interaction is considered 
zero in nature. We 
consider these values as it has a dominant component in the contribution
of the hadronized phase. We can thus compute the total modified free energy $F_{total}$ as,
\begin{equation}
 F_{total}=\sum_{i} F_{i}~+~F_{interface}~+~F_{h},
\end{equation}
~ where $i$ stands for $u$,~$d$
and $s$~quark and gluon.
\section{Thermodynamic properties and speed of sound}
\par The thermodynamic properties like pressure, entropy and specific
heat of the system can be calculated from the total
free energy. The standard thermodynamics give the following
relations~\cite{rajiv}:
\begin{equation}
 Pressure~P=-(\frac{\partial F}{\partial v})
\end{equation}
\begin{equation}
Entropy~ S= -(\frac{\partial F}{\partial T})
\end{equation}

\begin{equation}
Specific~heat~ C_{v}= T(\frac{\partial S}{\partial T})_{v}
\end{equation}
The behavior of entropy and specific
heat with temperature indicate the nature of
phase transition of the system.
So, we calculate the entropy and specific heat for a flow 
parameter $\gamma_{q}=1/8$ and gluon parameter
$ \gamma_{g} =8 \gamma$. At this particular value we
can get the stability of the droplet with maximum radius of $r=4.2~fm$. 
The stability is found more
with increasing gluon flow parameter up to $\gamma_{g}=10 \gamma_{q}$. With
the increasing gluon parameter the pressure of the system is increased and
very much effective in the free energy expansion, Yet 
the  entropy and specific heat is not effective with the increasing gluon
parameter. Now using these entropy and specific heat
we calculate the speed of sound of QGP.
The speed of sound is given as the ratio of these two thermodynamic 
parameters~\cite{ali}:
\begin{equation}
Speed~of~Sound~ C_{s}^2 = \frac{S}{C_{v}}
\end{equation}
The speed of sound is found to be almost constancy over the range of 
temperature and the model parameter. 
\begin{figure}[h]
\centering
\includegraphics[width=7cm,clip]{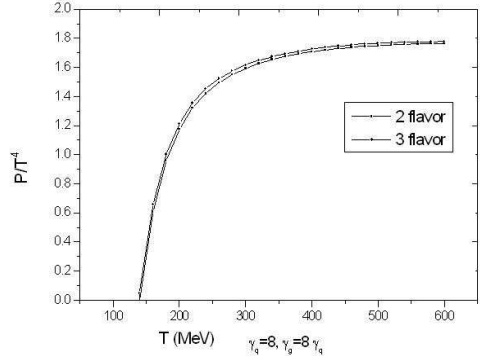}
\caption{The variation of $P/T^{4}$~vs. T at~$\gamma_{q}=1/8~$, $\gamma_{g}=8\gamma_{q}$
at the critical radius.}
\label{fig-1}       
\end{figure}
\begin{figure*}[htb]
\centering
\includegraphics[width=7cm,clip]{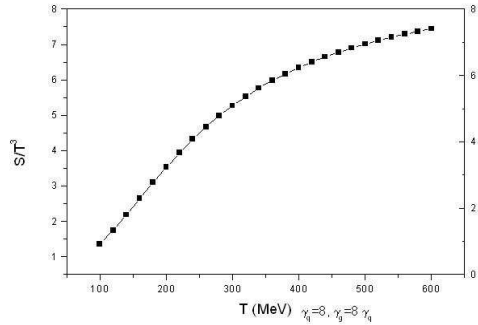}
\caption{Entropy~$S/T^{3}$ vs.~T~at~$\gamma_{q}=1/8~$, $\gamma_{g}=8\gamma_{q}$ at the critical radius.}
\label{fig-2}       
\end{figure*}
\begin{figure}[htb]
\centering
\includegraphics[width=7cm,clip]{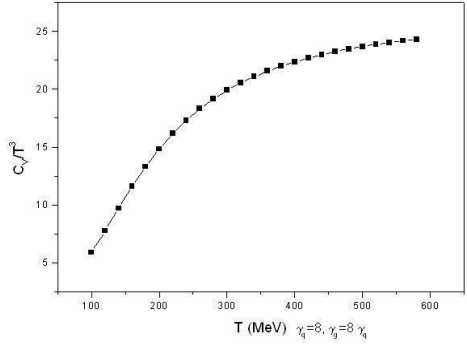}
\caption{ Specific heat~$C_{v}/T^{3}$ vs.~T~
at~$\gamma_{q}=1/8~$, $\gamma_{g}=8\gamma_{q}$ at the critical radius.}
\label{fig-3}       
\end{figure}
\begin{figure}[htb]
\centering
\includegraphics[width=7cm,clip]{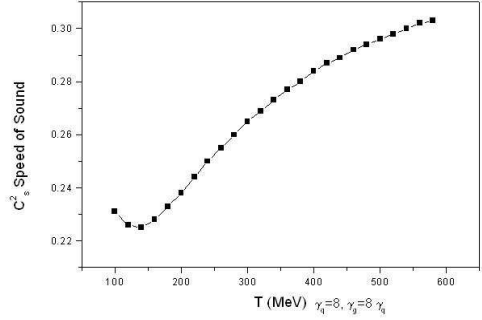}
\caption{Speed of sound~$C_{s}^2$ vs.~T~
at~$\gamma_{q}=1/8~$, $\gamma_{g}=8\gamma_{q}$ at the critical radius.}
\label{fig-5}       
\end{figure}


\section{Results:} The thermodynamic properties like pressure, entropy and 
specific heat are calculated for quark and gluon 
flow parameter with the introduction of one loop
correction factor in the interacting mean-field potential. 
They are calculated for a particular chosen flow parameter which fit well
in the evolution of free energies with the stable largest droplet size.
Free energy expansion can be produced with different stable
droplet size with increasing gluon flow parameter
up to $\gamma_{g}=10 \gamma_{q}$. There is change in the pressure
but the parameter does not affect in
the entropy and specific heat.
 The outputs from the calculation are shown in the figures.
 The figure~$1$ shows the variation of $P/T^{4}$ with the temperature~$T$ 
for three flavor and two flavor of quarks.
The variation in the figure determines the equation of state (EOS) of 
the system.
The equation of state is an essential component to understand the nature
of QGP and to model the behavior of QGP in early universe formation.
The variation agrees with the recent outcomes of 
lattice results with insignificant difference. In Fig.$2$ we
show the plot of $S/T^{3}$ with the temperature. The
plot indicates the variation at lower temperature. It again shows
the constancy
over higher temperature. The result almost follows the lattice predictions
for three and two flavors of quarks and it produced similar results
with our earlier works without the loop correction~\cite{dhar}. 
Again Fig. $3$ shows specific 
heat changes with
the temperature. The specific heat is almost independent with
higher temperature and the plot matches with the recent calculation of
specific heat. Moreover, it is 
indicated by lattice QCD that formation 
of QGP droplets
take place under the condition of roll-over phase rather than a sharp
jump as temperature varies~\cite{aoki}. 
From the two thermodynamic parameters we 
use to calculate the speed of sound of the system. The result is
shown in the figure $~4~$. 
From the figure, the speed of sound is well behaved
in nature in terms of the lattice results of sound speed
 and almost the same to our earlier
calculation of speed of sound without the loop correction. 
The result is approximately
close to $C_{s}^{2}=0.3$ which is obtained with this range of temperature.
It is quite similar to the latest value of speed of sound of 
lattice QCD~\cite{fr}.

\section{conclusion:}
We conclude that due to the introduction of one
loop correction in the mean field potential, we can describe 
the EOS through pressure and 
behavior of entropy and specific heat. So, we further
obtained the speed of sound from these two thermodynamic
parameters as $C_{s}^2=0.3$ which is independent
from the model parameter and temperature. It is a good 
result in comparison with recent lattice results.
\subsection{\bf Acknowledgments:}

The authors thank K K Gupta for useful discussions
and critical reading of the manuscript. The work is supported by  
the University of Delhi, Delhi, India providing research 
and development grants.


\begin{thebibliography}{}
\bibitem{karsch}
E. Karsch, A. Peikert and E. Laermann, Nucl. Phys. B {\bf 605}, 579 (2001).
\bibitem{csernai}
L. P. Csernai, J. I. kapusta, E. Osnes, Phys. Rev. D {\bf 67}, 045003 (2003);
A. Tawfik, Phys. Rev. D {\bf 71}, 054502 (2005). 
\bibitem{satz}
H. Satz,{ \it CERN-TH-2590}, 18pp (1978); F. Karsch, E. Laermann, A. Peikert, Ch. Schmidt and S. Stickan, Nucl. Phys. B {\bf 94}, 411 (2001).
\bibitem{mustafa}
M. G. Mustafa, D. K. Srivastava and B. Sinha, Euro. Phys. J. C {\bf 5}, 711 (1998);  F. Karsch and H. Satz, Nucl. Phys. A {\bf 702}, 373 (2002).
\bibitem{peshier}
A. Peshier, B. K$\ddot{a}$mpfer, O. P. Pavlenko and G. Soff,
Phys. Lett. B {\bf 337}, 235 (1994); V. Goloviznin and H. Satz, Z. Phys. C {\bf 57}, 671 (1993).
\bibitem{raman}
R. Ramanathan, K. K. Gupta, A. K. Jha and S. S. Singh, Pram. J. Phys. {\bf 68}, 757 (2007).
\bibitem{singh}
S. S. Singh, D. S. Gosain, Y. Kumar and A. K. Jha, Pram. J. Phys. {\bf 74}, 27 (2010).
\bibitem{somoren} 
S. S. Singh, K. K. Gupta and A. K. Jha, Int. J. Mod. Phys. A {\bf 29}, 1450094 (2014);
S. S. Singh and R. Ramanathan, Prog. Theo. Exp. Phys. {\bf 103Do2}, 2014.
\bibitem{ramanathan}
R. Ramanathan, Y. Mathur, K. K. Gupta and A. K. Jha, Phys. Rev. C {\bf 70}, 027903 (2004).
\bibitem{fermi}
E. Fermi, Zeit F. Physik {\bf 48}, 73 (1928); L. H. Thomas , Proc. Camb. Phil. Soc. {\bf 23}, 542 (1927); H. A. Bethe, Rev. Mod. Phys. {\bf 9}, 69 (1937).
\bibitem{brambilla}
 N. Brambilla, A. Pineda, J. Soto and A. Vairo, Phys. Rev D {\bf 63}, 014023 (2001).
\bibitem{melnikov}
K. Melnikov and A. Yelkhovsky, Nucl. Phys. B {\bf 528}, 59 (1998);
A. H. Hoang, Phys. Rev D {\bf 59}, 014039 (1999).
\bibitem{fischler}
W. Fischler, Nucl. Phys. B {\bf 129}, 157 (1977); A. Billoire, Phys. Lett B {\bf 92}, 343 (1980).
\bibitem{smirnov}
A. V. Smirnov, V. A. Smirnov, M. Steinhauser,
Phys. Lett. B {\bf 668}, 293 (2008);
A. V. Smirnov, V. A. Smirnov and M. Steinhauser,
Phys. Rev. Lett. {\bf 104}, 112002 (2010).
\bibitem{neergaad}
G. Neergaad and J. Madsen, Phys. Rev. D {\bf 60}, 05404 (1999);
M. B. Christiansen and J. Madsen, J. Phys. G {\bf 23}, 2039 (1997).
\bibitem{mardor}
I. Mardor and B. Svetitsky, Phys. Rev. D {\bf 44}, 878 (1991);
H. T. Elze and W. Greiner, Phys. Lett. B {\bf 179}, 385 (1986).
\bibitem{weyl}
H. Weyl, {\it Nachr. Akad. Wiss Gottingen} 110 (1911).
\bibitem{balian}
R. Balian and C. Block, Ann. Phys. (NY) {\bf 60}, 401 (1970).
\bibitem{rajiv}
R. V. Gavai, S. Gupta and S. Mukherjee, Pram. J. Phys. {\bf 71}, 487 (2008);
S. Bors\'{a}nyi, Nucl. Phys. A {\bf 904-05}, 270c (2013).  
\bibitem{ali}
A. A. Khan, $et~al$, Phys. Rev. D {\bf 64}, 074510 (2001);
Y. Aoki, Z Fodor, S. D. Katz and K. K. Szab\'{o}, J. High Energy Phys. {\bf 01}, 089 (2006).
\bibitem {dhar}
D. S. Gosain, S. S. Singh, Int. J. Theo. Phys. {\bf 53}, 2688 (2014).
\bibitem{aoki}
Y. Aoki, G. Endr\"{o}di, Z Fodor, S. D. Katz and K. K. Szab\'{o}, Nature {\bf 443}, 675 (2006).
\bibitem{fr}
F. Karsch, Nucl. Phys. A {\bf 783}, 13 (2007); H. B. Meyer, Euro. Phys. J. A {\bf 47}, 86 (2011). 
\end{thebibliography}
\end{document}